\documentclass[aps,preprint,nofootinbib,showpacs]{revtex4}
\usepackage{xcolor}
\usepackage[hypertex]{hyperref}
\usepackage{graphicx,subfigure}
\usepackage{amsmath,amsfonts,amssymb}

\newcommand{\f}{\frac}

\newcommand{\m}{\mathbf}

\graphicspath{{plots/}}
\begin{document}
\title{Surprising properties of water on its binodal  as the reflection of the specificity of intermolecular interactions}
\author{V.L. Kulinskii}
\altaffiliation[On leave from the ]{Department of Theoretical Physics, Odessa National University, Dvoryanskaya 2, 65026 Odessa, Ukraine}
\affiliation{Department of Molecular Physics, Kiev National
University, Glushko 6, 65026 Kiev, Ukraine}
\email{kulinskij@onu.edu.ua}
\author{N.P. Malomuzh}
\email{mnp@normaplus.com}
\affiliation{Department for Theoretical Physics, Odessa National University, Dvoryanskaya 2, 65026 Odessa,
Ukraine}
\begin{abstract}
In the paper the behavior of density (or specific volume), the
heat of evaporation and entropy per molecule for normal and
heavy water on their coexistence curves is discussed. The
special attention is paid on the physical nature of the
similarity in the behavior of density and the heat of
evaporation for water and argon as well as the nearest water
homolog $H_{2} S$. It is shown that the appearance of this
similarity is a consequence of the rotational motion of water
molecules, which averages the inter-particle potential in water and leads it to the argon-like form. To describe the fine
distinctions in the behavior of the binodals for water, $H_{2} S$ and argon the dependence of the proper molecular volume on pressure is taken into account. In accordance with this often used the van der Waals and Carnahan-Starling equations of states are modified. The very surprising behavior of the entropy diameter for water is analyzed. It is shown that the nontrivial details of the temperature dependence for the entropy diameter is directly connected with the peculiarities of the rotational motion of molecules in water. The effect of strong dimerization of water molecules in the fluctuation region near the critical point is studied in details.
\end{abstract}
\maketitle
\section{Introduction}\label{sec_intro}
\begin{figure}
\centering
\includegraphics[scale=0.8]{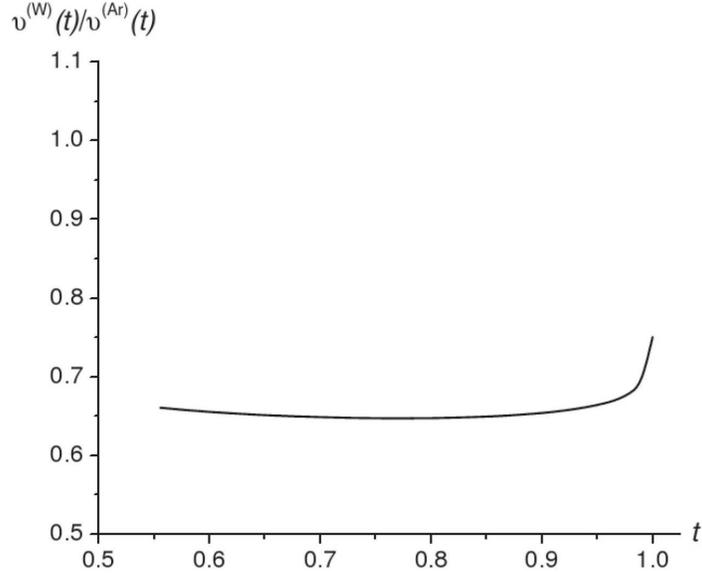}
\caption{Temperature dependence of the ratio
${R_{v}  = v^{(w)}(t) / v^{(Ar)}(t)}$ on the coexistence
curves of water and argon according to
\cite{water_argonlike_maloleinik_cpl2008}.}\label{fig1_ratio}
\end{figure}

The surprising properties of water, as it is well known
\cite{book_franks_water1,book_eiskauzman_water}, are caused by
H-bonds. The character of their spatial organization and time
dynamics depends essentially on temperature and pressure. As a
result, the peculiarities of the translational and rotational
motion of water molecules are very different near the triple
and critical points. There are several key facts which allow to
make the important conclusions about the behavior of the H-bond
network in water and the character of the thermal motion of
water molecules, in particular, near its critical point. One of
them is presented in Fig.~\ref{fig1_ratio}, where the
comparison of the specific volume per molecule $v^{(i)}$ for
water and argon (i = w; Ar) is given. The dimensionless
temperature $t = T/T_c(i)$ , where $T_c(i)$ is the critical
temperature of liquids, is used. As seen, the behavior of the
specific volume of water is very surprising. Practically in the
whole region it is argon-like. Only in the narrow vicinity of
the critical point ($0.95 < t < 1$) the deviation of
$v^{(w)}(t)$ from the argon-like dependence is essential.
Outside of this region ($t_m < t < 0.95$), where $t_m$ = 0.42
is the melting point, the deviation from the argon-like
dependence does not exceed $3 \div 4\%$ . The minimum of the
specific volume of water at $T \approx 277K$ corresponds to the
weak increase of the ratio $R_{\upsilon}  (t) = v^{(w)}(t) /
v^{(Ar)}(t)$ at approaching the melting point. In accordance
with Fig.~\ref{fig1_ratio} the temperature dependence of
$R_{v}(t)$ can be represented in the form:
\[R_{v}(t) = \kappa + r_{H} (t),\]
where $r_{H} (t)$ is the contribution caused by H-bonds in water: $r_{H} (t)\ll \kappa $.

To describe the temperature dependence of $r_{H} (t)$ seems to
be natural to apply to the Hilbert's principle
\cite{water_argonlike_maloleinik_cpl2008,water_lokmalzak_jstructkhim2003},
which was formulated for the first time in the algebraic
invariant theory and having the numerous applications in the
statistical hydrodynamic \cite{book_moninyaglom_1}. According
to this principle, an arbitrary complicated function can be
expanded in the series with respect to independent primitive
functions, which have the same properties of symmetry:
\begin{equation}\label{rh_defin}
r_{H} (t) = {\sum\limits_{k} {r_{k}^{(H)} \cdot S_{k} (t,p)}}\,,
\end{equation}
where $r_{k}^{(H)} $ are the coefficients, which are functions
of temperature $t$ and the dimensionless pressure $p = P /
P_{c} $, $P_{c} $ is the critical pressure. The independent
structural characteristics of the H-bond network, so-called
structural functions \cite{water_lokmalzak_jstructkhim2003},
play a role of primitive functions ${\left\{ {S_{i}}\right\}}$.

The most important characteristics of the H-bond network are
the average number $n_H$ of H-bonds per molecule and the
parameter of tetrahedricity $\chi(T)$ (see
\cite{water_naberukhin_zhstructkhim1997,water_paschekgeiger_jphyschem1999}).
Therefore, the first two structural functions can be chosen as:
$S_{1} = n_{H} (t,p)$ and $S_{2} = \chi _{H} (t,p)$,. The
structural functions of a higher order are assumed to be
responsible for the finer details of the H-bond network and
here they will be ignored. In
\cite{water_argonlike_maloleinik_cpl2008} it was shown,
that with good accuracy \eqref{rh_defin} can approximated as
\begin{equation}
\label{rh_approx}
r_{H} (t) = r_{H}^{(0)} \cdot n_{H} (t), \quad n_{H} (t) = 4(1 - \lambda t + ...).
\end{equation}
Fitting the experimental data on $R_{v}(t)$ with the help of
\eqref{rh_approx} one can get: $\kappa = 0.63$, $r_{H}^{(0)} =
0.015$; $\lambda = $0.85. The value and the temperature
dependence of $n_{H} (t)$, given by Eq.~\eqref{rh_approx}, are
in good agreement with the results obtained in
\cite{water_naberukhin_zhstructkhim1989,water_dielspect_pccp2002,water_1coordsphere_science2004}.

Below we will show, that the temperature dependence of the heat
of evaporation in water is also similar to that for argon. The
estimate for $n_{H}(t)$, following from the analysis of the
heat of evaporation, practically coincides with
Eq.~\eqref{rh_approx}.

To explain the similarity in the behavior of the specific
volume and the heat of evaporation in water and argon we should
conclude that the mentioned quantities and some other
thermodynamic characteristics of water are determined by the
averaged intermolecular potential, which takes the argon-like
behavior.

From Eq.~\eqref{rh_approx} it follows that near the critical
point the average number of H-bonds per molecule takes value
$n_{H} (1) \approx 0.7$. It means, that in the fluctuation
region, $0.85 < t < 1$, water can be considered as the ensemble
of dimers. This conjecture is supported by the results obtained
in \cite{water_dimer_us_nato2007}. Besides, it becomes to be
self-evident if we take into account that in this region the
volume of cavity ($v_{cav}= 2\,v_{m}$) occupied by two water
molecules are greater than that volume, which is created by a
rotating dimer ($v_{cav} > v_d$). Qualitatively, almost full
dimerization of water molecules allows us to explain naturally
the sharp enough increment of the ratio $R_{v}$ near the
critical point (see Fig.~\ref{fig1_ratio}).

The present paper is devoted to the careful analysis of the
density, entropy and the heat of evaporation on the coexistence
curve of water in the wide temperature range from the melting
to the critical points. The special attention is paid to the
dimerization effects of water molecules in the fluctuation
region.

To describe the fine details in the behavior of the density,
diameter of entropy and the heat of evaporation on the binodal
we take into account the rotation of water molecules. Due to
this the proper molecular volume becomes to be dependent on the
temperature and density. The last circumstance is proved to be
essential not only for water but also for all liquids with
non-spherical molecules and even for the noble gases, although
in smaller extent.

The comparative analysis of thermodynamic properties of the
normal and heavy water is performed. It notices that the
isotopic effect manifests itself not only in the shifts of the
critical and melting points. In the considerably more extent it
is manifested in the behavior of their diameters of entropy and
their specific heats.

It is taken into account that in the fluctuation region the
volume, occupied by two water molecules, is greater than that
volume, which is created by a rotating dimer. As a result the
strong dimerization in water is stimulated. Using the methods
of chemical equilibrium, we study the degree of dimerization.
Applying the van der Waals equation of state, we show that the
location of the critical point in water is mainly determined by
the ensemble of dimers. It is established that different values
of the specific heats for normal and heavy water in their
fluctuation regions, $0.85 < t < 0.98$, are connected with the
inner rotation of monomers, entering the dimers $(D_{2} O)_{2}
$ in the heavy water.
\section{The dimerization of water in the near critical region}\label{sec_assoc}
In this section in accordance with
\cite{water_dimer_us_nato2007} we use the vdW based model with
the account of dimerization. We calculate the average number of
$H$-bonds per molecule along the coexistence curve. Note that
for completely dimerized water $n_{H} = 1$. In accordance with
said in Introduction we assume that water near critical point
is fully dimerized. Thus water dimer $(H_2O)_2$ plays the role
of a ``particle``. According to the quantum chemistry
calculations \cite{water_dimermc_jcp2007} the diameter of a
dimer $\sigma_d \approx 3.5 \AA$ and the dipole moment
$d_{d}\approx 2.6 \,D$.

The idea that the dimers are predominant in near critical
region has been put forward before (see
\cite{water_dimers_raman_jcp1998} and references therein).
Though the inference was made from the analysis of structural
data (Raman scattering) which did not led to the firm
conclusion. Lately, the results of MD simulations
\cite{water_dimermc_jcp2007} favored the picture that in near
critical region water behaves more like associative fluid as
opposed to the assumptions about peculiar (fractal) structure
of hydrogen bonded network
\cite{water_factalhbond_jcp2005,water_factalhbond_oleinikova_pccp2007}.

Therefore we can expect that in near critical region water can
be reasonably described by the vdW EOS for the system of
dimers. The basis for the use the van der Waals like EOS in
near critical region is due to low density and thus intensive
rotation. Indeed, the average spacing between dimers at the
critical density is about $3.5\,\AA$ and is more than the size
of dimer. However at removing of the critical point along
liquid branch the character of the dimer rotation changes: the
quasi-free rotation of a dimer discontinues at $v_{cav} \to
v_{d}$. When $v_{cav}(t) \le v_{d}$, the dimers breaks and
linear molecular chains are formed. A molecule entering in some
chain can change its spatial orientation in consequence of two
mechanisms: 1) the rotation around two H-bonds connecting it
with other molecules in this chain and 2) the break of these
H-bonds.  passes to the rotation around the H-bond connecting
dimer molecules. At the further removing of the critical point
linear molecular chains begin to cooperate in spatial clusters.
Here a molecule can change its orientation only because of the
break of H-bonds connecting it with the nearest neighbors.

\begin{figure}[hbt!]
\subfigure[\,]{\includegraphics[scale=0.56]{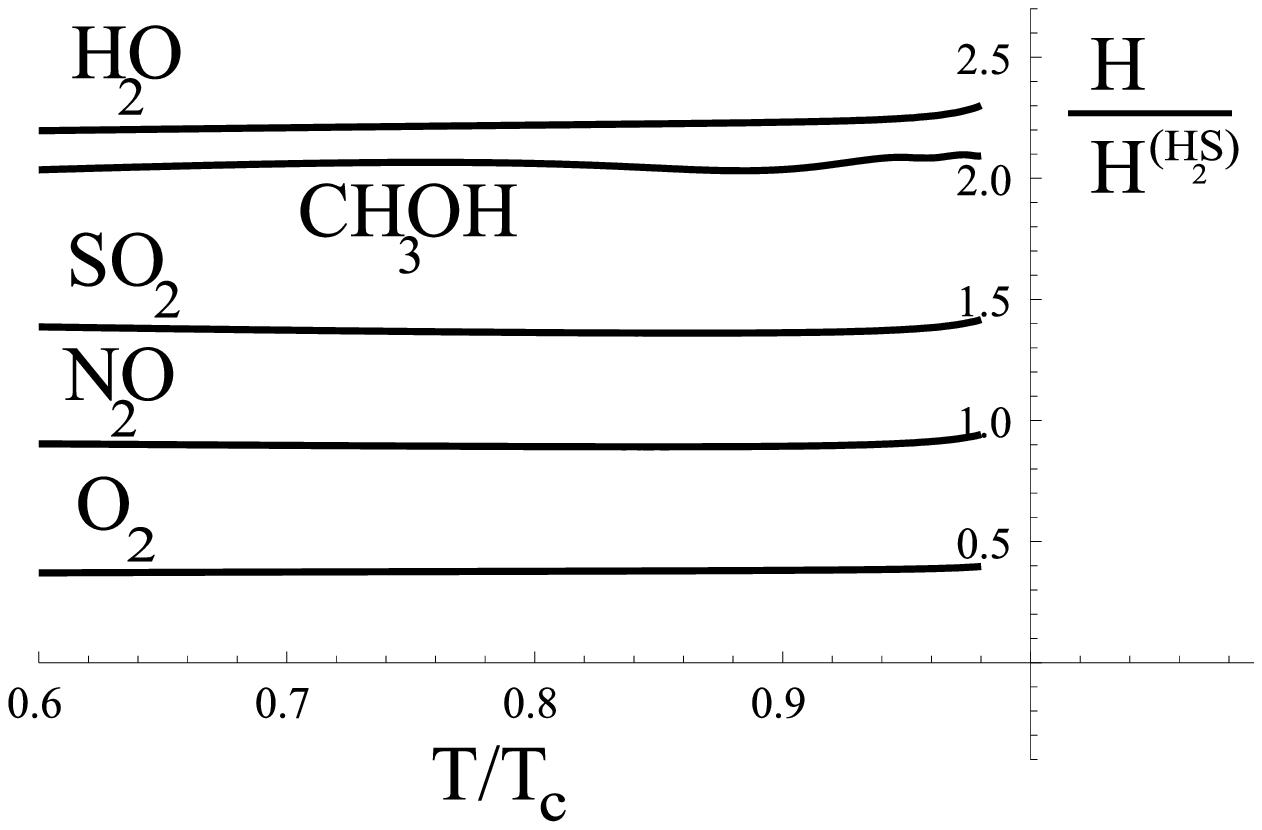}}
\hspace{0.5cm} \subfigure[\,]{\includegraphics[scale=0.6]{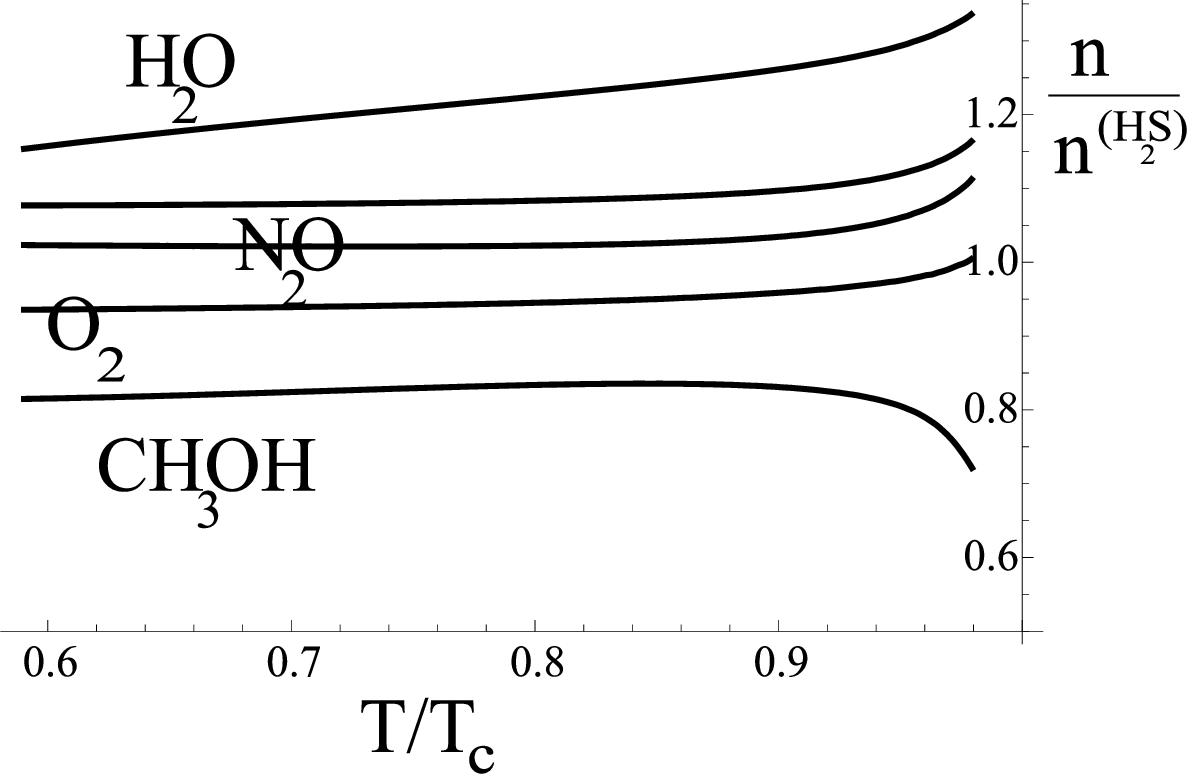}}
  \caption{The temperature dependencies of the ratios
  $R_{H}(i)$ (a) and $R_{n} (i)$ (b) according to data of
  \cite{nist69}}\label{fig_evaporheat_rel}
\end{figure}

The comparison of the heat of evaporation data gives the
additional support for the picture of rotating molecular
dipoles. It is known that the heat of evaporation of water is
very big in comparison with other homologs like $H_2S$ or
nonpolar simple liquids like $Ar$. This ratio for a number of
the liquids is shown on Fig.~\ref{fig_evaporheat_rel} with
$H_2S$ chosen as the reference fluid. Note that water and
methanol have the same ratio.

Note that the ratio of the heat of evaporation per molecule is
almost constant for different fluids along the whole
coexistence curve. The same constancy of the ratio of specific
volume takes place for water and argon (see
\cite{water_argonlike_maloleinik_cpl2008}). This fact
seems to be quite surprising because the interparticle
interaction differs for different liquids. As we see, the ratio
of the densities for water and its nearest homologue linearly
increases within 10\%. This increment is slightly larger than
for corresponding ratio for water and argon (see Inroduction
and \cite{water_argonlike_maloleinik_cpl2008}). This
fact means that the rotational motion of the weakly asymmetric
molecules of liquid   influences its EOS. This effect, as it
follows from Fig.~\ref{fig_evaporheat_rel}(b) has similar
character for other liquids. Below it will be connected with
the softness of the molecular volume entering the vdW EOS  and
close ones.

To explain the constancy of the ratio $R_{H}(i) =
\f{H^{(i)}}{H^{(H_2S)}}$ as well as $R_{n}(i) =
\f{n^{(i)}}{n^{(H_2S)}}$ we are forced to suggest that the
simplest thermodynamic quantities are determined by the
averaged inter-particle potentials. In fact this averaged
potential appear to be of the van der Waals character. At that
time, such quantities as heat capacity, the density, the
entropy diameters and dielectric permittivity are determined by
the deviation from the averaged interactions. These quantities
strongly depend on the short range correlations.

To explain the numerical values of the ratio $R_{H}(i)$  for
different liquids, we can use the following qualitative
reasons. By order of magnitude, the heat of evaporation is the
work per molecule which is necessary to expand its specific
volume from $v_{l}$ to $v_{g}$: $H^{(i)}\approx P\,\left(\,
v_{l}-v_{g}\,\right)\,,$ where all quantities are taking on the
coexistence curve. Outside the fluctuation region $v_{l} >
(\gg) v_{g}$, therefore we can approximately write
$H^{(i)}\approx P\,v_{g}$ or $H^{(i)}\approx
Z^{(i)}_{c}\,T^{(i)}_c\,\tilde{Z}^{(i)}$, where $\tilde{Z}$ is
the dimensionless form for the compressibility factor $Z =
\f{P\,v}{T}$. From here it follows that $R_{H}$ is estimated
as:
\begin{equation}\label{evheat_ratio}
  R_{H}(i)  = \f{T^{(i)}_c}{T^{(H_2S)}_c}\,\f{Z^{(i)}_c}{Z^{(H_2S)}_c}\,
  \f{\tilde{Z}^{(i)}}{\tilde{Z}^{(H_2S)}}\,,
\end{equation}
For the rare enough vapor phases of $i$-th  and $H_2S$ system
\[\f{\tilde{Z}^{(i)}}{\tilde{Z}^{(H_2S)}}\approx 1\,,\] so:
\begin{equation}\label{evheat_ratio_estimate}
  R_{H}(i) \approx \f{T^{(i)}_c}{T^{(H_2S)}_c}\,\f{Z^{(i)}_c}{Z^{(H_2S)}_c}\,.
\end{equation}
\begin{table}
\begin{tabular}{|c|c|c|c|c|c|c|}
\hline
  Fluid&$CH_3OH$&$H_2O$&$O_2$&$Ar$&$N_2O$&$SO_2$\\
  \hline
 $R_{H}^{(m)}(i)$, Eq.~\eqref{evheat_ratio_estimate}&$0.9$&$1.4$& $0.42$& $0.41$&$0.8$&$1.1$\\
\hline
$R_{H}^{(exp)}(i)$&$2.0$&$2.4$& $0.37$& $0.35$&$0.9$&$1.4$\\
\hline
\end{tabular}
\caption{The comparison of $R_{H} (i)$ calculated according to
Eq.~\eqref{evheat_ratio_estimate} for ensemble of monomers
(subscript $(m)$) and with the help of experimental data
\cite{nist69}.}\label{tab_evapheatratio}
\end{table}

As is seen from Table~\ref{tab_evapheatratio} the raw estimate
Eq.~\eqref{evheat_ratio_estimate} gives unsatisfactory
prediction for liquids like water and methanol. Like water, the
last is known of its associative properties
\cite{dimers_methanol_jpc1970,dimers_methanol_jpc1971}. Note
that the compressibility factor is calculated assuming that
only the monomers is present in near critical region. The value
of $R_{H}(i)$ for water with account of dimerization degree is
$2.5$ (see Appendix~\ref{sec_app_chemeq} and
Eq.~\eqref{zc_assoc}). Thus it allows to conclude that the
dimerization for the methanol as well as water is essential in
near critical region (see below).

From our consideration it follows that the determination of the
energy of H-bond for water from the heat of evaporation data is
not correct. In accordance with our estimates, in the vicinity
of the critical point water can be considered as the ensemble
of dimers with small admixture of monomers. To determine the
location of the critical point for water we apply to the van
der Waals EOS:
\begin{equation}
\label{vdweos}
p(n,T) = {\frac{{nT}}{{1 - nb}}} - an^{2}
\end{equation}
where $p$ is the pressure and $a$ and $b$ are the vdW
parameters. According to \cite{book_ll5} they are:
\begin{equation}\label{vdwpar_ab}
b= \f{2\pi}{3}\,\sigma_d^3\,,\quad a=\pi\,\int_{\sigma_{d}}^{\infty}\,U(r)\,r^2\,dr\,,
\end{equation}
In
\cite{crit_coulomb_crition_kulimalo_pre2003,water_dimer_us_nato2007}
it had been shown that the attractive part of the dimer-dimer
potential is mainly determined by their dipole-dipole
interactions:
\begin{equation}\label{u_dipdip}
U_{a}(r) = \frac{2}{3}\frac{1}{T}{\frac{d_{\dim} ^{4}}{r^{6}}}.
\end{equation}
Therefore the coefficient $a$ becomes to be temperature
dependent: $a \to \frac{a}{T}$, and the van der Waals EOS
transforms to Bertlo EOS:
\begin{equation}\label{bertlo_eos}
 p(n,T) = {\frac{{nT}}{{1 - nb}}} - {\frac{{a}}{T}}n^{2}\,.
\end{equation}
The coordinates of the critical point is determined by the standard conditions:
\[
\frac{\partial\,p}{\partial\, n} = 0\,,\quad
\frac{\partial^{\,2}\,p}{\partial\, n^{\,2}} = 0\,.\] The
coordinates of the critical point for the Bertlo
EOS~\eqref{bertlo_eos} are determined by equations:
\begin{equation}\label{vdwcp_coord}
  n^{(d)}_{c} = \f{1}{3b_d} =  \f{1}{2\pi\,\sigma_{d}^3}\,,\quad
  T^{(d)}_c = \sqrt{\f{8a_{d}}{27b_d}} = \f{2\lambda^2\sqrt{2}}{9}\,\f{e^2}{\sigma_d}\,,\quad p^{(d)}_c=\f{3}{8}\,n_c\,T_c = \f{\lambda^2}{12\pi\sqrt{2}}\,\f{e^2}{\sigma_d^4}\,,
\end{equation}
where $\lambda$ is the parameter which determines the dipole
moment $d_{dim}$ of the dimer and $d_{dim} ~= ~\lambda
e\,\sigma_d $. We choose the value of $\lambda$ so that to get
the best fit for the coordinates of the vdW critical point
which are:
\begin{equation}\label{bestfit}
T_{c}^{(d)} \approx 647\, K\,,\quad P_{c}^{(d)} \approx 18.0\,
MPa\,,\quad \rho_{c}^{(d)}\approx 322\, kg/m^3\,.
\end{equation}
The fitting gives the value  $\lambda \approx 0.19$ and the
dipole moment of a dimer is $d_{dim} \approx 2.9 D$, which is
in good agreement with the value for dimer dipole moment
$d_{dim} = 2.6 D$ \cite{water_dimerdipole_jcp2004}. Sure total
dimerization is an approximation but above estimates show that
it is consistent. Thus the difference in the compressibility
factor $Z_c$ is due to the deviation from vdW EOS since the
influence of free monomers was neglected.

Now we refine the estimate for the location of the critical
point taking into account that water near the critical point is
the mixture of dimers and small quantity of monomers. Let us
introduce the degree of the dimerization as following:
\begin{equation}\label{dmrz}
  n_{1} = (1-A)\,n_0 \qquad n_{2} = \f{A}{2}\,n_0\,,
\end{equation}
where $n_0=n_1+2n_2$ is the number density  of water molecules
both free and dimerized, $n = n_1+n_2$ is the number density of
particles (both the dimers and the monomers). The coefficients
$a$ and $b$ are the vdW parameters \cite{book_hansenmcdonald}:
\begin{equation}\label{bertlo_ab}
b = c_1b_1+c_2b_2\,,\quad a = c_1^2\,a_{11} + 2\,c_1c_2 a_{12}+c_2^2\,a_{22}\,,
\end{equation}
where
\begin{equation}\label{vdwpar}
b_i= \f{2\pi}{3}\,\sigma_i^3\,,\quad a_{ij}=
\pi\,\int_{\f{\sigma_{i}+\sigma_j}{2}}^{\infty}\,U_{ij}(r)\,r^2\,dr\,,
\end{equation}
and $c_{i}= \f{n_i}{n_1+n_2}$ are the concentrations of the
monomers and dimers correspondingly with $n_i$ being the number
density of $i-$th species respectively, $\sigma_{i}$ are the
diameters of the particles and $U_{ij}$ is the potential of the
interaction between particles of i-th and j-th kinds. In
\cite{water_dimer_us_nato2007} it was shown that the
interaction between monomers and dimers is determined mainly by
the dipole-dipole forces. The corresponding potentials have the
form:
\[U_{ij} = \beta\,\f{2}{3}
 \f{
\left\langle\, \m{d}^2_i \,\right\rangle\left\langle\, \m{d}^2_j \,\right\rangle
 }{r^6}\]
In general the virial coefficient $a_{ij}$ is the sum of the
vdW term which does not depend on the temperature and the
temperature dependent dipole-dipole term. The chemical
potential for monomers is:
\[\mu_1 =  -T\ln  \left( 1-n\,b
 \right) +\frac {T n b_1}{1-nb}-2n\left(\,c_1\,a_{11}+c_{2} a_{12}\,\right)
\]
The chemical potential of the dimers is obtained by the change
$1\to 2$. To construct the binodal for the mixture of dimers
and monomers we use common condition of equilibrium the
equality of the chemical potentials and the pressures of the
coexisting phases:
\begin{equation}\label{mup}
  \mu_{liq} = \mu_{vap}\,,\quad p_{liq} = p_{vap}\,.
\end{equation}
together with the chemical equilibrium equation
Eq.~\eqref{chemeq} to determine the equilibrium number
densities of the species in coexistence phases.


Let us discuss the compressibility factor as a function of the
degree of dimerization. For water considered as an ensemble of
monomers and described by the vdW EOS is expected to be
\begin{equation}\label{zcvdw_mon}
Z^{(vdW)}_m =3/8\,.
\end{equation}
If water is completely dimerized:
\begin{equation}\label{zcvdw_dim}
Z^{(vdW)}_d = 3/16\,.
\end{equation}
This value is close to the experimental one $Z^{(exp)} = 0.23$  near the critical point of water and it can be considered as one of evidences in the favor of high degree of dimerization near the CP. The values of $Z_c$ corresponding to $n_c = \rho_c/m_{a}$, where $\rho_c$ is the critical mass density and $m_{a}$ is the atomic mass, for different fluids considered as the monomeric ones are in Table~\ref{tab_zc}.

As seen from Table~\ref{tab_zc} the compressibility factor for methanol practically coincides with that estimate for water which is obtained for the ensemble of dimers. This fact also points to the high degree of dimerization in methanol near its critical point \cite{dimers_ammonia_assoc_jcp_1979,dimers_ammonia_water_jcompchem_1982}, that is natural since the intermolecular interaction in methanol is determined by H-bonds, similar to that in water. For several other liquids in Table~\ref{tab_zc} do not forming H-bonds the values of the compressibility factor near their critical points are noticeably higher than for water and methanol. This difference is probably connected with 1) the essentially less degree of dimerization in them, even for Ar (see \cite{crit_dimers_noblepcs_physica2009}), and 2) the approximate character of the vdW EOS.

\begin{figure}[hbt!]
  \includegraphics[scale=0.5]{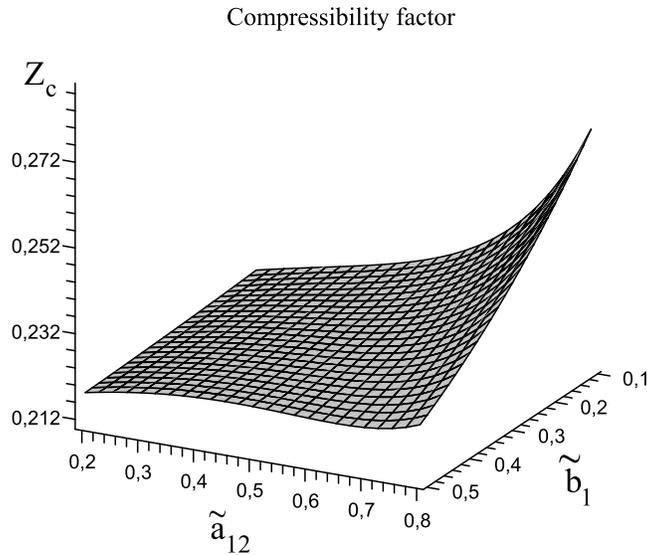}\\
  \caption{Compressibility factor as the function of dimensionless volume of the monomer $\tilde{b}_1$ and dimensionless monomer-dimer virial coefficient $\tilde{a}_{12}$ (see Eq.~\eqref{bertlo_tilde_ab}) at the energy of dissociation $E_d/T_c = 3.5$.}\label{fig_zc}
\end{figure}
The influence of the interparticle interactions on the behavior of the compressibility factor is presented in Fig.~\ref{fig_zc}. As for the locus of the CP, the most strong dependence is observed on $\tilde{a}_{12}$.

Ending this Section let us briefly discuss the interaction between molecules
outside the fluctuation region. Here, as was noted above, the equation of
state for water is determined by the averaged inter-particle potential.
Besides the hard-core part it includes the contributions of the dispersive
forces, the dipole-dipole and H-bond interactions between monomers:
\[\left\langle\, U(1,2) \,\right\rangle \approx U_{hc} (r_{12} ) + U_{dis} (r_{12} ) + U_{mm} (r_{12} )+ U_{H} (r_{12} ).\]
The term $U_{mm} (r_{12} )$ is similar to Eq.~\eqref{u_dipdip}:
\[U_{mm} (r) = {\frac{2}{3}}{\frac{1}{T}}{\frac{{d_{mon}^{4}
}}{r^{6}}}, \]
but $\left( {{\frac{{d_{mon}}} {{d_{\dim}} } }} \right)^{4} < 0.1$.
Therefore we conclude that outside the fluctuation region $U_{mm} (r_{12} )\ll U_{dis} (r_{12} )$, while inside it $U_{dis} (r_{12} ) \ll U_{a} (r)$. The term $U_{H} (r_{12} )$ plays the noticeable role only on small distances between monomers. In fact, it leads to some renormalization of the hard-core diameter. Due to this we expect that the behavior of the specific volume per molecule, the heat of evaporation and some other thermodynamic quantities outside the fluctuation region are determined by the potential:
\[ \left\langle\, U(1,2) \,\right\rangle  \approx U_{hc} (r_{12} ) + U_{dis} (r_{12} ).\]
The vdW EOS \eqref{vdweos} with the coefficient $a$ do not depending on temperature can serve as quite satisfactory zero approximation for the EOS of water.
%

\begin{table}
\begin{tabular}{|c|c|c|c|c|c|}
\hline
  Fluid&$CH_3OH$&$H_2O$&$H_2S$&$Ar$&$C_6H_6$\\
  \hline
$Z_c$&$0.19$&$\m{0.23}$&$0.28$&$0.29$&$0.27$\\
\hline
\end{tabular}
\caption{The compressibility factor $Z_c$ for fluids.}\label{tab_zc}
\end{table}
\section{The rotation of molecules and the behavior of the entropy diameter.}\label{sec_rotationanddiameter}
\begin{figure}
\centering
  \subfigure[\,entropy diameter]{\includegraphics[scale=0.65]{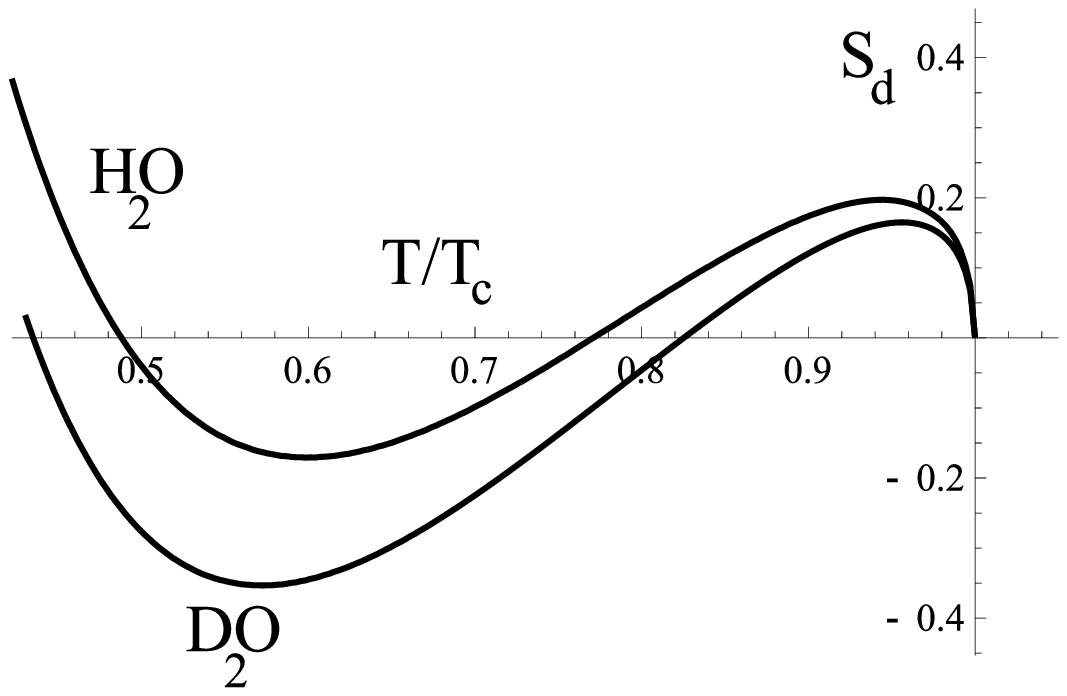}
  \label{fig_sdiam_h2od2o}}\hspace{0.5cm}
\subfigure[\,density diameter]{\includegraphics[scale=0.6]{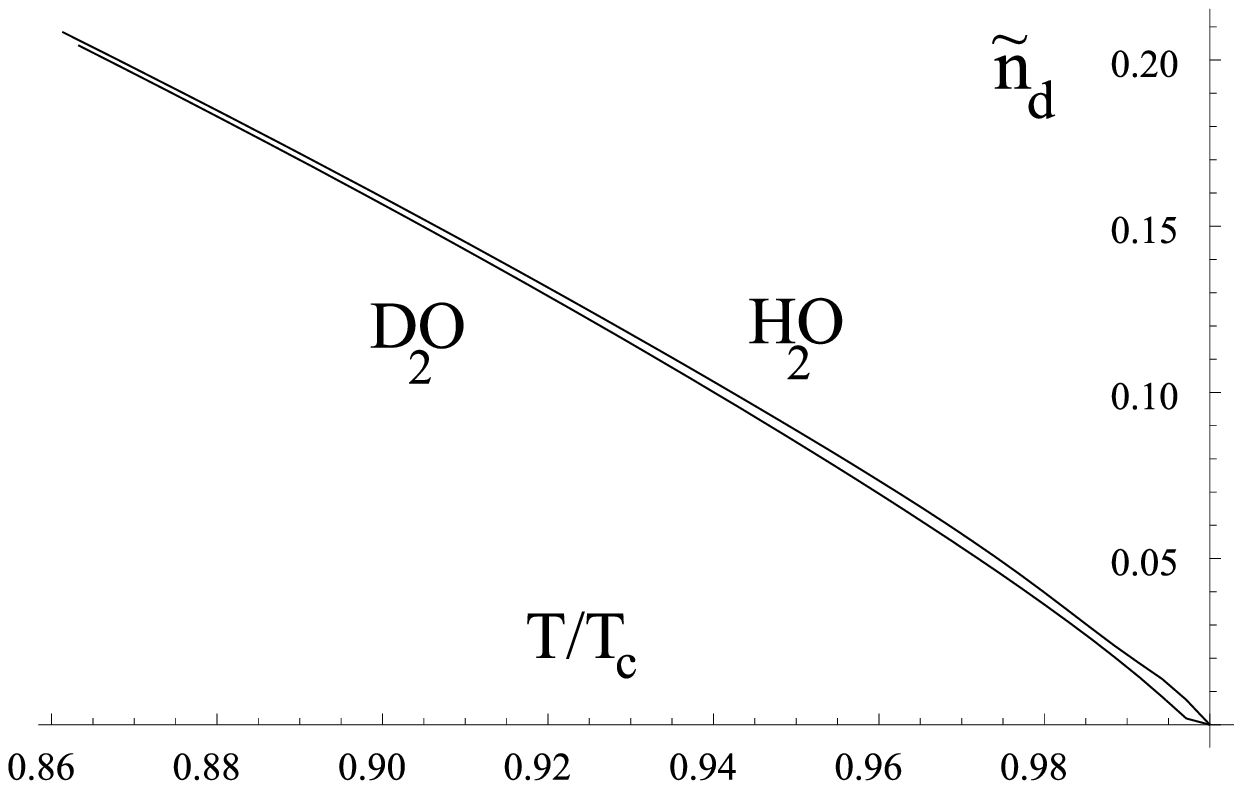}\label{fig_ddiam_h2od2o}}
  \caption{The diameters of the entropy (a) and the density (b) for $H_2O$ and $D_2O$ according to experimental data \cite{nist69}.}\label{fig_diam_h2od2o}
\end{figure}

The character of the molecular motion differs in liquid and gaseous phases. The number density (or specific volume) is the natural order parameter for the liquid-vapor critical point. However, it is one-particle characteristics of a system and therefore it describes only crude enough behavior of a system.  From this point of view the choice of the entropy as the order parameter seems to be more informative since the entropy is determined by correlation effects of all orders \cite{book_ziman_disorder}. Besides the part, reducible to density, this order parameter includes also the contributions reflecting the more fine details of the molecular motion. Such discrepancy is reflected in the asymmetry of the binodal. The degree of the asymmetry of the coexistence curve is described by the diameter of density:
\begin{equation}\label{diameter}
  \tilde{n}_{d} = \f{1}{2}\,\left(\,\tilde{n}_{l}+\tilde{n}_{g}\,\right) - 1\,,
\end{equation}
where $\tilde{n}_{i} = n_{i}/n_{c}\,, i= l, g$\,, $n_{l}, n_{g}, n_{c}$ are the number densities of the liquid, gas branches of the binodal and at the critical point correspondingly. The analogous quantity can be defined for the entropy. The temperature dependence of the entropy diameter:
\begin{equation}\label{entrop_diam}
  S_{d} = \f{S_{l}+S_{g}}{2} - S_c
\end{equation}
for normal and heavy waters is presented in Fig.~\ref{fig_diam_h2od2o}. The difference in the values of $S_d$ is caused by the isotopic effect (see also Section~\ref{sec_isotop}).  The comparison of $S_{d}$ for water with those for $H_2S, O_2$ and $Ar$, which has the spherical molecules, shows that non-monotone temperature dependence of the entropy diameter arises only for non-spherical molecules for which the rotational degrees of freedom should be taken into account.

\begin{figure}
\centering
\includegraphics[scale=0.8]{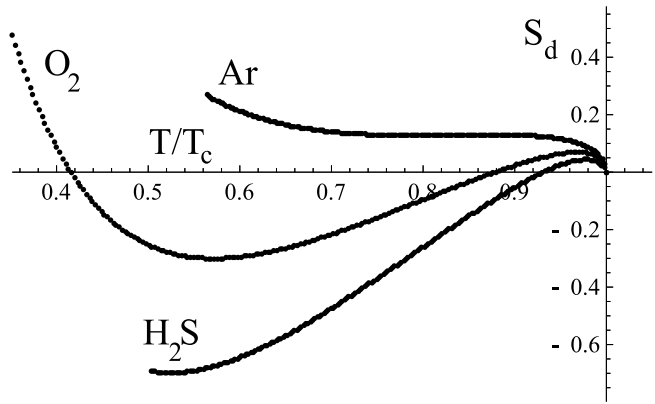}
  \caption{The diameters of the entropy for different liquids according to experimental data \cite{nist69}.}\label{fig_sdiam_repr}
\end{figure}
For all systems with non-spherical molecules there are two characteristic temperatures $T_{l}$ and $T_{u}$  at which $S_{d} = 0$. In particular, for normal and heavy waters $T_{l}$ and $T_{u}$ take the following values (see Fig.~\ref{fig_sdiam_h2od2o}):
\begin{align*}
  H_2O\,: &\quad T_{l} = 316 \, K\,,\quad  T_{u} = 499 \, K\,,\\
  D_2O\,: & \quad T_{l} = 280 \, K\,,\quad  T_{u} = 531 \, K\,.
\end{align*}

The nontrivial temperature dependence of $S_{d} (t)$ for water and substances of type $H_{2} S$ and $O_{2} $ is connected with the change of the rotational motion of molecules when density of a system increases. In normal and heavy waters H-bonds play additional important role. If the average number of H-bonds per molecule is close to three or more, a molecule can only oscillate about temporary equilibrium position. Let $\tau_{0} $ be the average time for such oscillations, so called the residence time. In \cite{water_bulmalpank_jschemrus2005,water_malfisenko_chemphys2008,water_bulmalok_jml2008} it was shown that the value of $\tau_{0} $ is essentially more than the characteristic time of the rotational motion $\tau_{r} \sim \left( {I / k_{B} T} \right)^{1 / 2}$, where $I$ is the inertia moment, only for $T < T_{H} $, where $T_{H} \approx 315K$. It is very surprising that the diameter of entropy changes its sign namely at this temperature ($T_{l} = 316K)$. At $T > T_{H} $ the residence time of the local molecular configurations is comparable with $\tau _{r} $, therefore the rotational motion of water molecules will influence on the behavior of $S_{d} $ analogously to that in liquid $H_{2} S$ or $O_{2} $.

\begin{figure}
\includegraphics[scale=0.7]{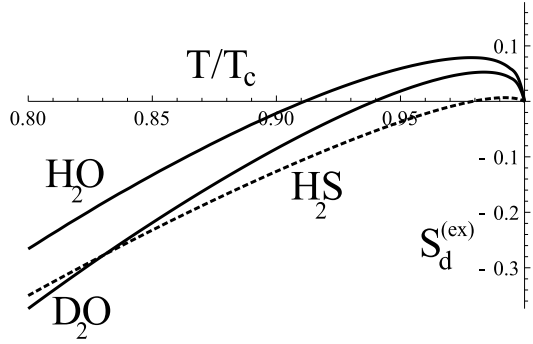}
  \caption{The diameter for residual (excess) entropy.}\label{fig_sdiam_excess}
\end{figure}

The character of the temperature dependence becomes partly clearer when we pass to the excess entropy
\begin{equation}\label{entropy_excess}
S^{(ex)} = S - S^{(id)},
\end{equation}
where
\begin{equation}\label{entropy_ideal}
S^{(id)}(T) = c_{\upsilon}^{(id)} \ln {\frac{T}{{T_{c}}} } - \ln {\frac{{n}}{{n_{c}}} }, \quad c_{\upsilon}^{(id)} = {\frac{{k}}{2}},
\end{equation}
is the entropy of ideal gas, $k$ is the number of thermalized degrees of freedom which are treated classically. As it follows from Fig.~\ref{fig_sdiam_excess} the most essential difference between the behavior of $S_{d} $ for normal and heavy water as well as $H_{2} S$ is observed in the fluctuation region, where the both kind of water are close to be fully dimerized state.

As has been said above the behavior of the diameter of the entropy reflects the change of the rotational motions of the molecules with the density . One can see that (heavy)water differs in this respect from other homologs. The situation looks like there is some deficit (see Fig.~\ref{fig_sdiam_excess}) of the order in liquid phase of water especially near the critical point. From the point of of the dimerization picture such a fact acquires natural explanation. The interaction between water molecules is converted into formation of bound states, the dimers, which rotate freely in liquid phase in lower density near critical region.
Note that the effect of lowering the entropy of the liquid phase due to nonspherical shape of the repulsive molecular cores was investigated by  \cite{crit_corenonspher_chemphys1995} within generalized vdW  EOS model in low pressure region. Here we show that the entropy diameter is the fine characteristic which is very sensitive to such effect and can be described within the model EOS where the softness of the particle core is taken into account (see Section~\ref{sec_softcore}).
To explain the behavior of the diameter of the entropy we use the basic thermodynamical representation of the entropy in dimensionless units (in units of the Boltzmann constant $k_{B}$):
\begin{equation}\label{entrop_basic}
  S = S_c+ c_{v}\ln\f{T}{T_c} + f(n)-f(n_c)\,,
\end{equation}
where $S_c$ is the entropy at the CP, $c_v$ is the dimensionless specific heat, $f(n)$ is some function describing the density dependence of the entropy. From Eq.~\eqref{entrop_basic} one can conclude that:
\begin{equation}\label{sdiam_basic}
  S_{d} = \f{c^{(l)}_{v}+c^{(g)}_v}{2}\ln\f{T}{T_c} +\f{f(n_l)+f(n_g)}{2} - f(n_c)\,.
\end{equation}
As it follows from Eq.~\eqref{sdiam_basic}, the behavior of the diameter essentially depends on the specific form of the equation of state, which determines the function $f(n)$. In its turn, the EOS is sensitive to the effective proper molecular volume corresponding to the rotating non-spherical molecules in liquid and vapor phases. The detailed analysis of this question will be given in the following Section.

%

\begin{figure}
\centering
  \subfigure[\,]{\includegraphics[scale=0.6]{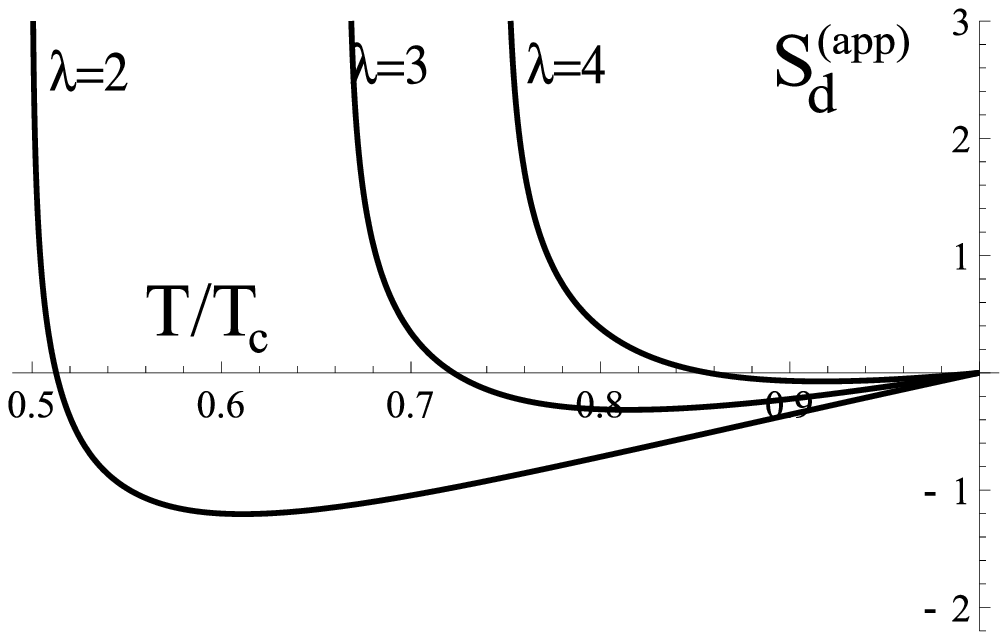}}\hspace{0.5cm}
  \subfigure[\,]{\includegraphics[scale=0.6]{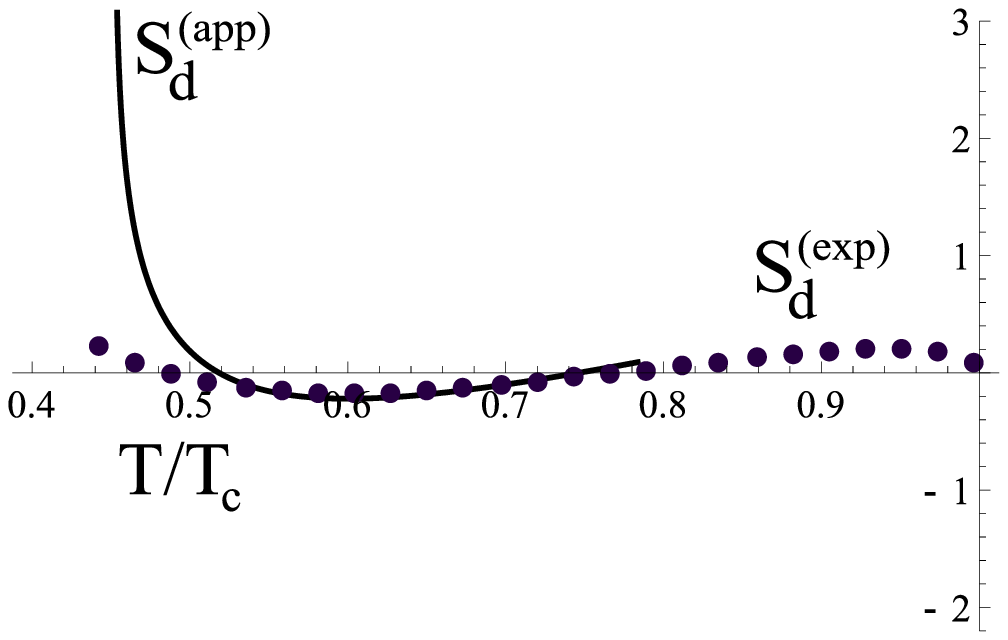}}
  \caption{$S^{(app)}_{d}$ according to Eq.~\eqref{sdiam_id} as a function of $\lambda$ (a), (b) the comparison of Eq.~\eqref{sdiam_id} (solid line) with the parameters $\Delta c_{v} = 5$ and $k=6$, $a\approx 0.73\,,b\approx 2.3 \,, \lambda \approx 3$ corresponding for water with the data (dots) for $S_d$ \cite{nist69}.}\label{fig_sdiam_id}
\end{figure}

Here we restrict ourselves only by the consideration of $S_{d}(T)$, for which $f(n)$ is taking in the ideal gas approximation. The branches of the binodal are taken in the mean field approximation:
\begin{equation}\label{densitydiff_mf}
\tilde {n}_{l} - \tilde {n}_{g} = 2b\vert \tau \vert^{1 / 2} + o(\tau )\,, \quad
\tau = {\frac{{T - T_{c}}} {{T_{c}}} },
\end{equation}
\begin{equation}\label{densitydiameter_mf}
\tilde {n}_{d} = a\vert \tau \vert + o(\tau ).
\end{equation}
From here it follows that
\begin{equation}\label{densitybinodal_mf}
\tilde {n}_{l,g} = 1\pm b\vert \tau \vert^{1 / 2} + a\vert \tau \vert + \ldots
\end{equation}
Supposing that outside the fluctuation region the specific heat capacities of liquid and gas phases take the constant values $c^{(g)}_v = \f{k}{2}$ and $c^{(l)}_v = c^{(g)}_v + \Delta c_{v}$, from Eqs.~\eqref{sdiam_basic} and \eqref{densitybinodal_mf} we get the following approximate expression for $S_{d}$:
\begin{equation}\label{sdiam_id}
 S^{(app)}_{d} (T) = \left(\,\f{\Delta c_{v}}{2}+ c^{(id)}_V\,\right)\,\ln\f{T_l}{T_c} - \ln \left(\,1-\lambda \,\left(\,1-\f{T_{l}}{T_c} \,\right)\,\right)\,,
\end{equation}
As it follows from \cite{nist69} for $Ar$ the value of $\Delta c_{v}$ is 1, for $H_2S$  $\Delta c_{v}\approx 2$ and for water (both normal and heavy) $\Delta c_{v} \approx 5$. The parameter $\lambda$ depends on the specific EOS and \[\lambda  = b^2 - 2\,a \,.\]
For example, for the vdW EOS $a=0.4, b=2$, for CS EOS $a=1.2, b\approx 2.5$. The behavior of $S^{(app)}_{d}(T)$ for different values of $\lambda $ is presented in Fig.~\ref{fig_sdiam_id}. Here it is necessary to note that $S^{(app)}_{d}(T)$ becomes to be the non-monotone function of temperature for $k > 3$, i.e. this peculiarity can be caused by both the rotational and vibration degrees of freedom. Of course, far away from the critical point, the contributions of the rotational degrees of freedom are determinative. However, already in a crude approximation described above we are able to estimate the position of the lesser root $T_{l} $ of the equation $S_{d} (T) = 0$. We see that despite the crudeness of the approximation Eq.~\eqref{densitybinodal_mf} for the binodal  at suitable values of $\lambda$ the reasonable values for the root $T_{l} $ are obtained.

In the Section~\ref{sec_softcore} it will be shown that the behavior of $S_{d} (T)$ is noticeably changed if $f(n)$ is taking in the Carnahan-Starling form with modified proper molecular volume.

We want to finish this Section by the brief discussion of physical cause which leads to the appearance of the upper root $T_{u} $ for the equation $S_{d} (T) = 0$. The definition Eq.~\eqref{entrop_diam} of the entropy diameter includes the value of entropy $S_{c} $ at the critical point. It is known that the value of $S_{c}^{(mf)} $ calculated in the mean field approximation (for example in the vdW or CS approximations) is higher than the experimental value $S_{c}^{(exp)} $. It is a result of the neglecting by the mode-mode interaction in the mean field approximation. Therefore for the correct comparison of $S_{d} (T)$ calculated on the basis of experimental data with its mean field analog $S_{d}^{(mf)} (T)$ it is necessary to pass to the modified entropy diameter $S_{d}^{(M)} = S_{d}^{(mf)} + \Delta S_{c} $, where $\Delta S_{c} = S_{c}^{(mf)} - S_{c} $. It is clear that the equation $S_{d}^{(M)} (T) = 0$ will take two roots: the value of $T_{l} $ is shifted and the additional root $T_{u} $ appears. This situation will be considered in more details in separated work.

\section{The ``soft`` core EOS model}\label{sec_softcore}
In this Section we discuss the nature of the non-monotone temperature dependence of the diameter of entropy. We will show that the region of negative values of $S_d$ is mainly caused by the variation of the molecular volume in dependence of pressure. The appearance of such dependence is natural for liquids with non-spherical molecules. Indeed, the proper molecular volume for them is determined as the volume of  cavity formed by arbitrary rotating molecule. If the rotation of molecules becomes restricted, the corresponding cavity volume diminishes. Obviously, such an effect should take place with increasing of density. In water the considerable additional influence on the value of the effective molecular volume is generated by H-bonds. To describe the influence of the effective molecular volume on the behavior of the diameter of entropy, we start from some EOS, in which the dependence of the molecular volume on pressure is taken into account. Then we reconstruct the free energy and after this we find entropy in vapor and liquid phases and also the diameter of entropy.

\begin{figure}
\centering
\subfigure[\,CP locus]{\includegraphics[scale=0.6]{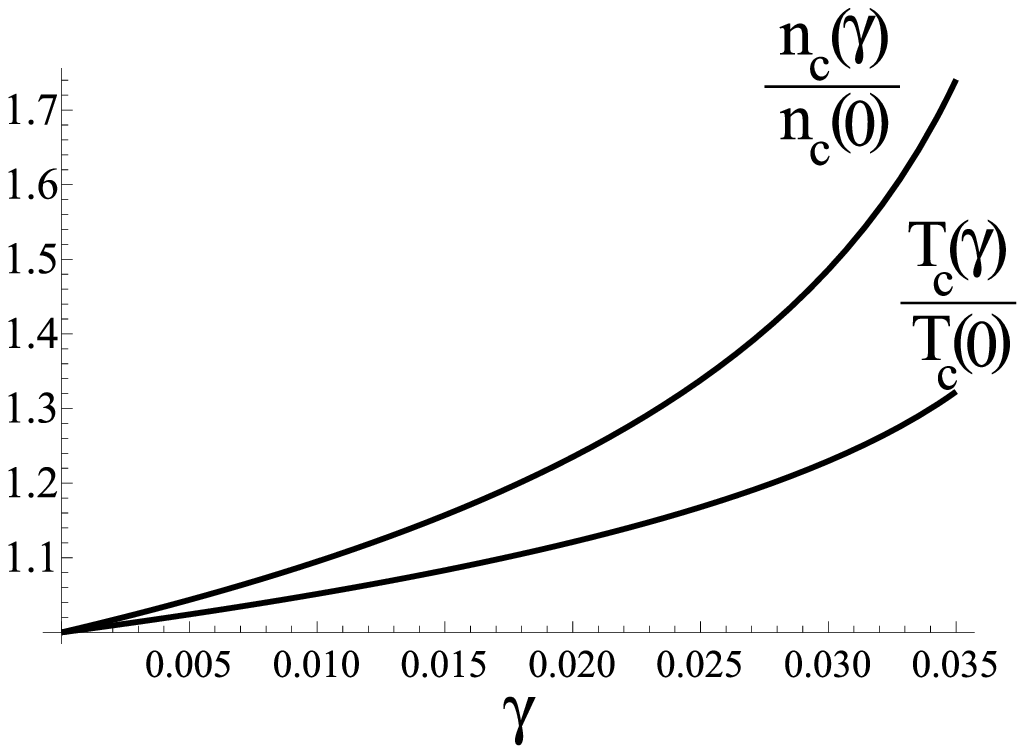}\label{fig_cplocus}}\hspace{0.5cm}
\subfigure[\,the binodal]{\includegraphics[scale=0.6]{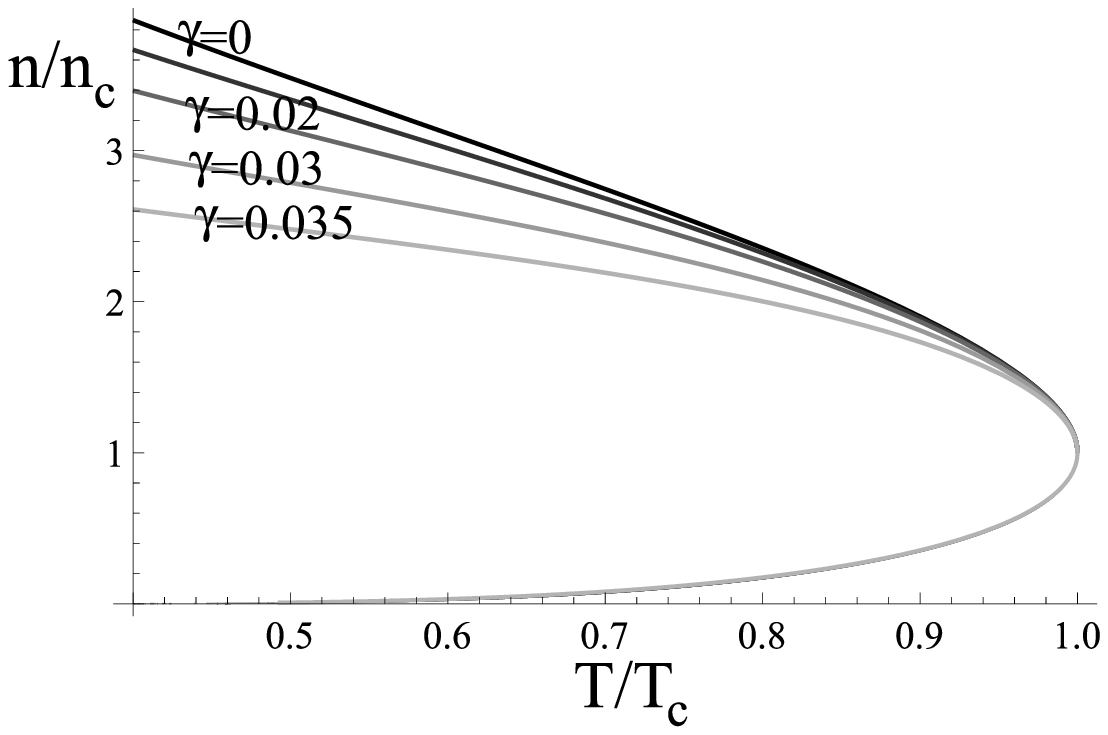}\label{fig_binodal}}
  \caption{The dependence of the coordinates of the CP (a) and the binodal (b) on $\gamma$ for the model Eq.~\eqref{free_hccs} with the soft core Eq.~\eqref{bmodel}.}
\end{figure}

\begin{figure}
\centering
\subfigure[\,monoatomic, $C^{(id)}_v = 3/2$]
{\includegraphics[scale=0.65]{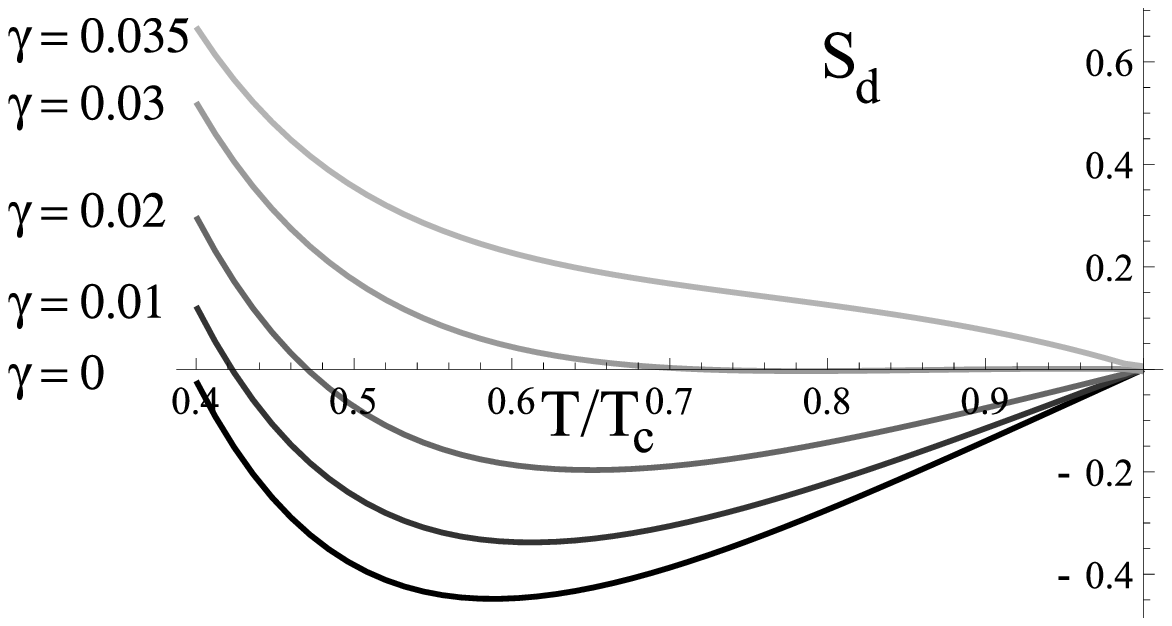}}\hspace{0.5cm}
\subfigure[\,diatomic, $C^{(id)}_v = 5/2$]
{\includegraphics[scale=0.65]{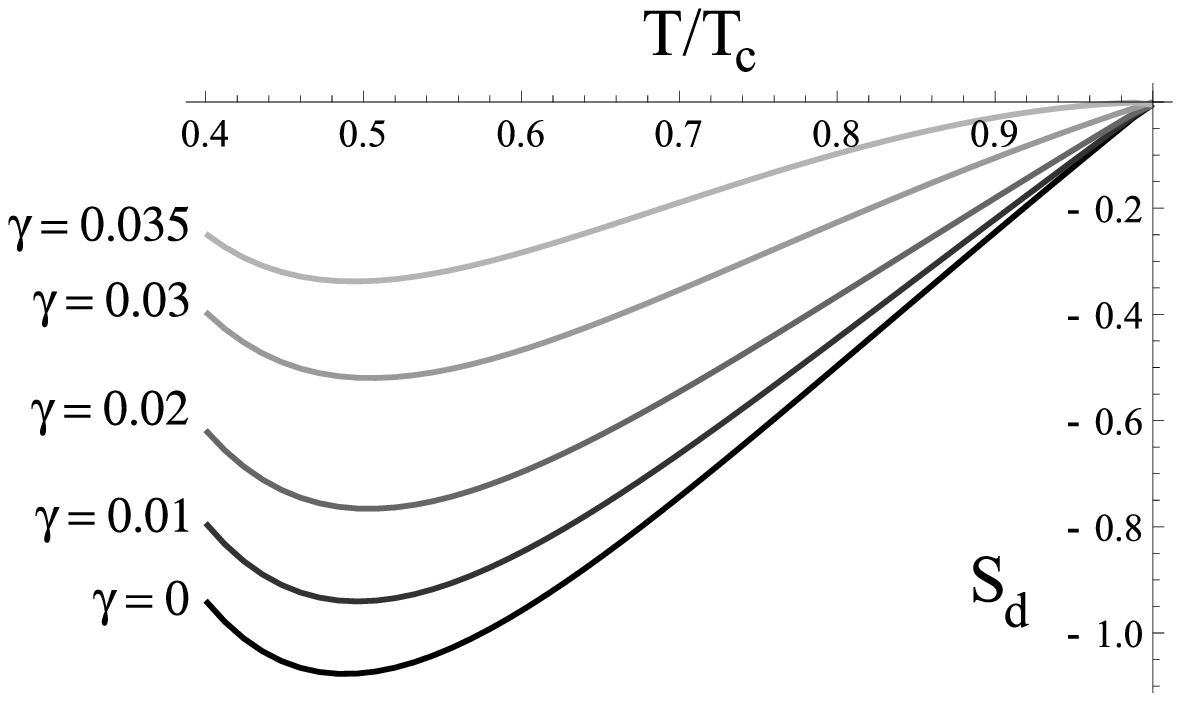}}
  \caption{The dependence of $S_d$ on $\gamma$ for Eq.~\eqref{free_hccs}.}\label{fig_sdiamodel}
\end{figure}

The EOS for simple molecular fluids is usually taken in the form \cite{eos_higginswidom_molphys1964,book_hansenmcdonald}:
\begin{equation}\label{peos}
  p = p_{+}(n,T)+p_{-}(n,T)\,,
\end{equation}
where $p_{+}$ is the pressure contribution due to hard core repulsive interactions and $p_{-}$ is the contribution of the attractive long range part of the potential. For simple fluids it is taken in the standard form:
\begin{equation}\label{p_attract}
p_{-} = -a\,n^2\,.
\end{equation}
From what has been said above it is clear that one can model the influence of the density on the rotation as the restriction of the angular configuration space available for the molecule in a cage formed by its neighbors. The change in the available space for free rotation of the molecule can be described by the dependence of the volume parameter $b$ on the density and the temperature. In accordance with said above we suppose that the density and temperature dependencies of the effective molecular volume $b$ can be modeled by the relation:
\begin{equation}\label{bmodel}
  b = \f{b_0}{1+\gamma \,p^{(id)}} \,,\quad p^{(id)} = \tilde{n}\,\tilde{T}\,.
\end{equation}
This parameter is included in the corresponding pressure term which describes the short range interaction (the hard core). The modified vdW EOS approximation for the free energy due to the interaction:
\begin{equation}\label{pgammavdw}
  \tilde{p} = \frac{\tilde{n}^2\, \tilde{T}}{(1+\gamma \,\tilde{n} \tilde{T})   (3(1+ \gamma \,\tilde{n} \tilde{T})-\tilde{n})}-\frac{9\,\tilde{n}^2}{8}+\tilde{n} \tilde{T}
\end{equation}
For small $\gamma$ the coordinates of the critical point are:
\[\f{T_c(\gamma)}{T_c(0)} = 1+\f{19}{9}\,\gamma+ o(\gamma)\,,\quad \f{n_c(\gamma)}{n_c(0)} = 1+3\,\gamma +o(\gamma)\,.\]
The analogous relations for EOS corresponding to the following model of the free energy:
\begin{equation}\label{free_hccs}
  F^{(CS)} = F_{id}+T\,\f{b\,n\left(\,4 - b\,n\,\right)}{\left(\,1-b\,n\,\right)^2} - a\,n\,,
\end{equation}
where $F_{id}$ is the ideal gas term are:
\[\f{T_c(\gamma)}{T_c(0)} = 1+t_1\,\gamma+ o(\gamma)\,, \quad \f{n_c(\gamma)}{n_c(0)} = 1+n_1\,\gamma +o(\gamma)\,,\qquad t_1 \approx 4.5\,,\,\,\,n_1\approx 8.1\]
The modification of the binodal for the Carnahan-Starling EOS at $\gamma \ne 0$ is shown on Fig.~\ref{fig_binodal}. As we see that the width of the binodal decreases with $\gamma$. This naturally reflects the diminishing of the phase asymmetry due to bigger compressibility of the effective volume, which in its turn means that the asymmetry between particle and the cavity (hole) decreases. From such point of view the most symmetrical is the binodal for liquid helium \cite{crit_dimers_noblepcs_physica2009}. The calculation of the entropy diameter $S_d$ supports this conclusion (see Fig.~\ref{fig_sdiamodel}).

Entropy of the system, corresponding to the Carnahan-Starling EOS, is determined in the standard way:
\begin{equation}\label{entropcs}
  S = -  \frac{\partial\, F^{(CS)}}{\partial\, T}
\end{equation}
The corresponding diameter of the entropy is presented in Fig.~\ref{subfig_sdiam_h2o}. Note that for water we neglected the temperature dependence of the coefficient $a$ in Eq.~\eqref{p_attract} in accordance with the discussion in Section~\ref{sec_rotationanddiameter}. Thus the EOS with the soft core forms the adequate basis for the successful description of the binodal  and fine nontrivial details in the behavior of the entropy diameter.

\begin{figure}
\centering
\subfigure[]{\includegraphics[scale=0.75]{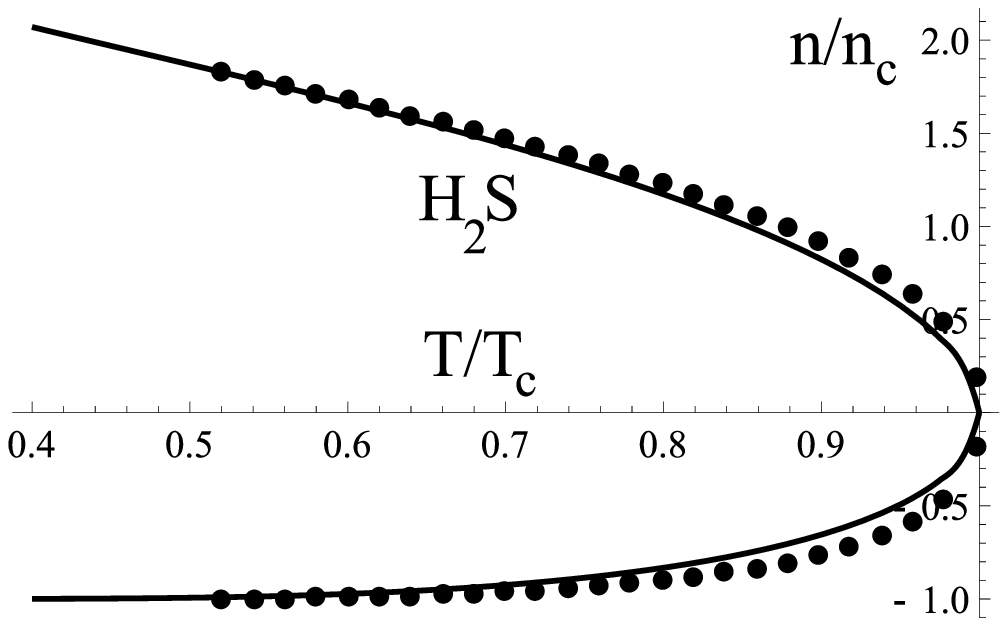}}\hspace{0.5cm}
\subfigure[]{\includegraphics[scale=0.75]{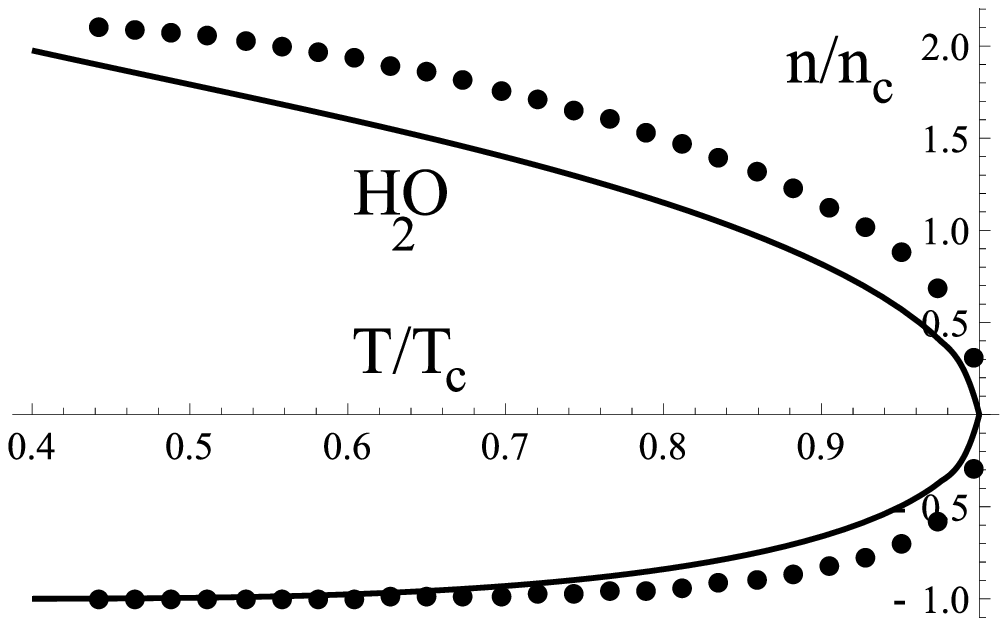}}
  \caption{The calculated binodal (solid) for the model Eq.~\eqref{free_hccs} with $\gamma \approx 0.03$  and the binodal for $H_2O$ and $H_2S$ (dots)}\label{fig_binodal_h2sh2o}
\end{figure}
\begin{figure}
\centering
\subfigure[]{\includegraphics[scale=0.6]{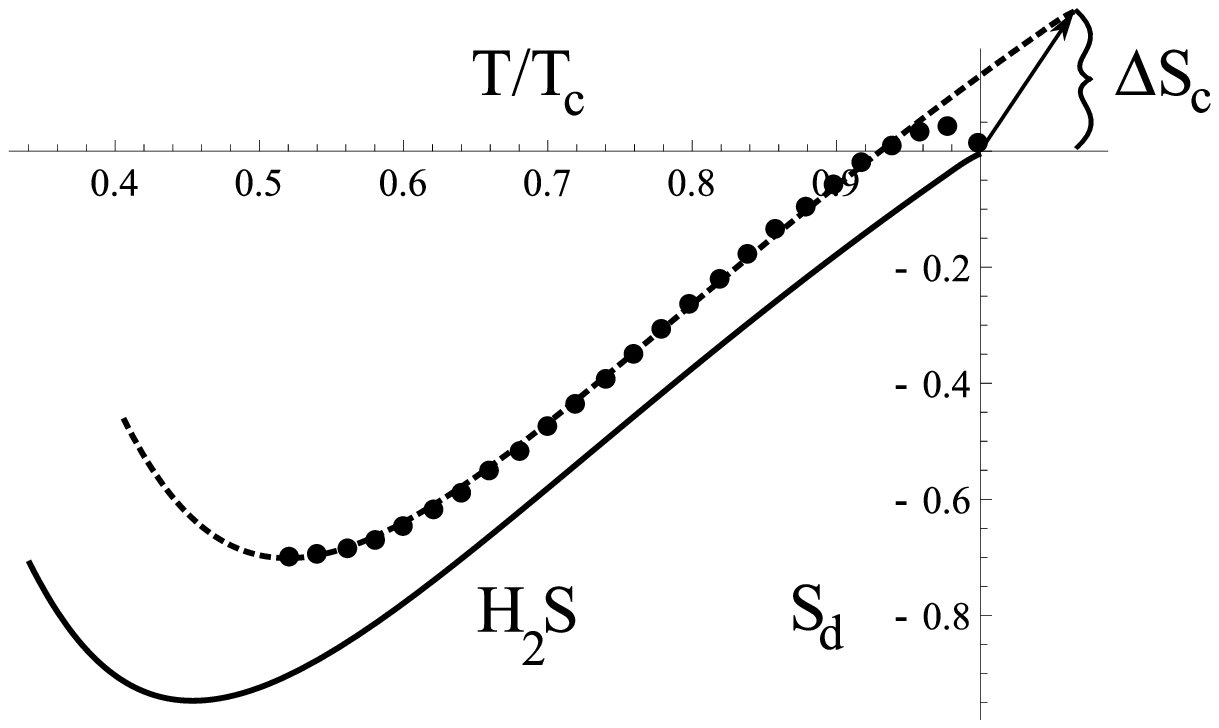}\label{subfig_sdiam_h2s}}\hspace{0.3cm}
\subfigure[]{\includegraphics[scale=0.6]{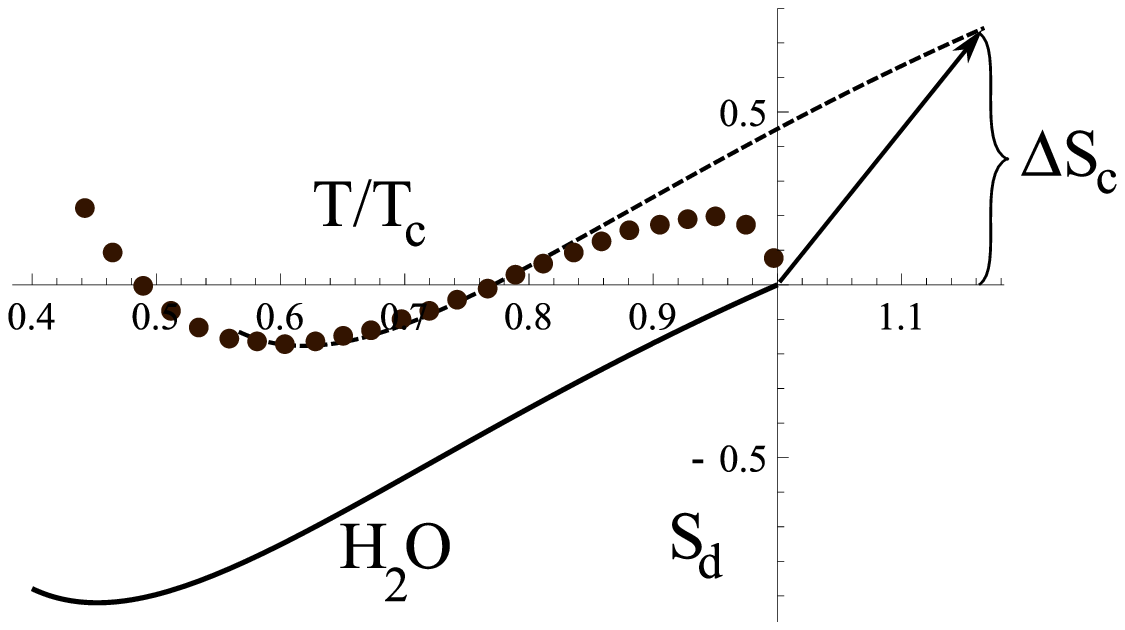}\label{subfig_sdiam_h2o}}
  \caption{The entropy diameter $S_d$ for the model Eq.~\eqref{free_hccs} with $\gamma \approx 0.03$ (solid line) and the data for $H_2O$ and $H_2S$ (dots). The shifted entopy diameter curve (see Section~\ref{sec_rotationanddiameter}) is shown by the dashed line.}\label{fig_sdiam_h2sh2o}
\end{figure}
\section{Manifestation of isotopic effect}\label{sec_isotop}
In this Section we want to complete our results presented above by the brief discussion of some distinctions in the behavior of the normal and heavy water. We focus our attention on the following characteristic manifestation of the isotopic effect: 1) small differences in the locations of the triple and critical points; 2) essentially more noticeable differences in the locations of the roots $t_{l} $ and $t_{u} $ of equations $S_{d} (t) = 0$ ($t_{k} = T_{k} / T_{c}^{(k)} $, $k = l,u)$ and 3) the nontrivial difference in the behavior of the specific heats. In these cases we observe different manifestations of the isotopic effect.

The critical temperature of normal water, $T_{c} (H_{2} O) = 647K$, is rather higher than one, $T_{c} (D_{2} O) = 644K$, for heavy water. It means, that the position of the critical point in heavy water is determined by the ensemble of dimers doped by small quantity of monomers, i.e. similarly to that for normal water. The small lowering of the critical temperature for heavy water is probably connected with 1) slightly different values of parameter $\lambda^{2} / \sigma _{d} $ for $H_{2} O$ and $D_{2} O$(see Eq.~\eqref{vdwcp_coord} and 2) some distinction of the dispersive interactions between molecules. Note, that in Eq.~\eqref{vdwcp_coord} the contribution of the dispersive forces is at all ignored.

The difference in the locations of the triple points is also insignificant: $T_{tr} (H_{2} O) = 273K$ and $T_{tr} (D_{2} O) = 277K$. From here it follows that normal and heavy water belong to the same class of corresponding states. For them $T_{tr} (H_{2} O) / T_{c} (H_{2} O) = 0.42$, $T_{tr} (D_{2} O) / T_{c} (D_{2} O) = 0.43$, that essentially differs from the ratios for the water homologue $H_{2} S$ and benzene: $T_{tr} (H_{2} S) / T_{c} (H_{2} S) = 0.50$, $T_{tr} (C_{6} H_{6} ) / T_{c} (C_{6} H_{6} ) = 0.50$. Such significant discrepancy is connected with the considerable influence of the dimerization on the location of the critical point. From Fig.~\ref{fig1_ratio} it follows that due to dimerization the specific volume per molecule increases approximately on 15\% . Assuming that the temperature shift has the analogous value: ${\frac{{\Delta T_{c}}} {{T_{c}}} }\sim {\frac{{\Delta v_{c}}} {{v _{c}}} }$, we obtain the following estimate for the critical temperature of non-dimerized water ($T_{c}^{(nd)} \approx T_{c} - \Delta T_{c} )$: $T_{c}^{(nd)} (H_{2} O) = 550K$. It leads to the estimate: $T_{tr} (H_{2} O) / T_{c}^{(nd)} (H_{2} O) \approx 0.50$. This fact can be considered as an indirect evidence for the dimerization near the critical point.

The locations of the roots $t_{l} $ and $t_{u} $ of equations $S_{d} (t) = 0$ for normal and heavy water satisfy the following relations:
\[ t_{l} (H_{2} O) - t_{l} (D_{2} O) \approx 0.053,\]
\[t_{u} (D_{2} O) - t_{u} (H_{2} O) \approx 0.054.\]
Their surprising closeness is not occasional. In the Section 4 it was stressed that the non-monotone temperature dependencies of the entropy diameter $S_{d}^{(k)} (t)$, $k = H_{2} O,D_{2} O$, is connected with the rotational motion of molecules. Therefore the comparison of the differences $\vert \Delta t_{l} \vert \approx \vert \Delta t_{u} \vert \approx 0.053$ with the characteristic parameter $\mu = {\frac{{\omega _{r} (H_{2} O) - \omega _{r} (D_{2} O)}}{{\omega _{r} (H_{2} O)}}}$ of the rotational motion is quite relevant. In principle, molecules can change their relative orientations by many ways; 1) each molecule rotates independently of others; 2) two nearest molecules change their orientations in concord, while all other neighbors retain in initial positions and so on. Note, that the dimerization in vapor phase of water leads to the natural lowering of entropy and negative values of $S_{d} (t)$. In liquid phase the existence of strong orientation correlations makes improbable the first way. At the same time, the corresponding pair in the second case can be identified with a dimer, for which we find an estimate:
\[\mu \sim 1 - \sqrt {{\frac{{m_{H_{2} O}}} {{m_{D_{2} O}}} }} \approx 0.05. \]
It is necessary to emphasize that we do not speak about the rotation of pairs (dimers) in liquid phase. Hence, the very close values of two dimensionless parameters is a strong argument in the favor of the second scenario of reorientations in vapor and liquid phases of normal and heavy water.

\begin{figure}
  \includegraphics[scale=0.75]{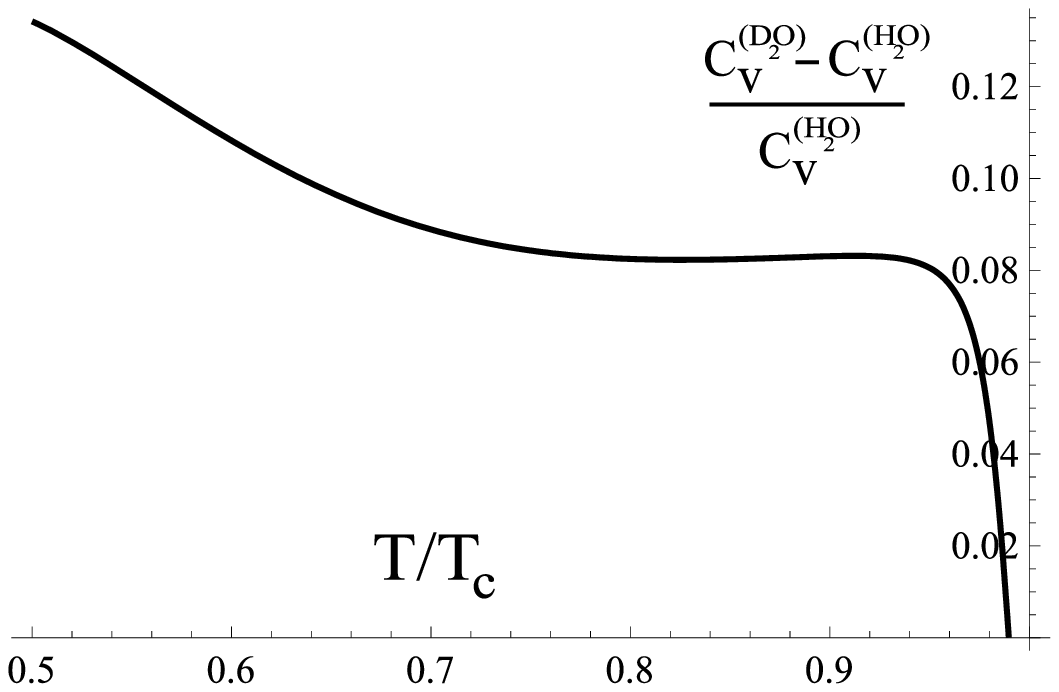}\\
  \caption{The relative value of the difference of the specific heats for normal and heavy water according to \cite{nist69}.}\label{fig_cvh2odo}
\end{figure}

\begin{figure}
  \centering
  \includegraphics[scale=0.75]{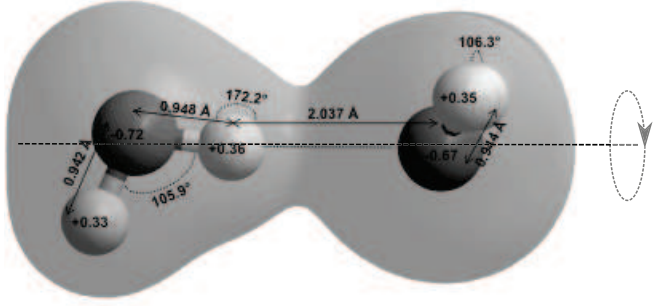}\\
  \caption{The relative rotation of the monomers entering a dimer. The parameters of water dimer are taken from \href{http://www1.lsbu.ac.uk//water//}{www1.lsbu.ac.uk/water/}}\label{fig_dimer_rotation}
\end{figure}

\begin{figure}
\centering
\includegraphics[scale=0.75]{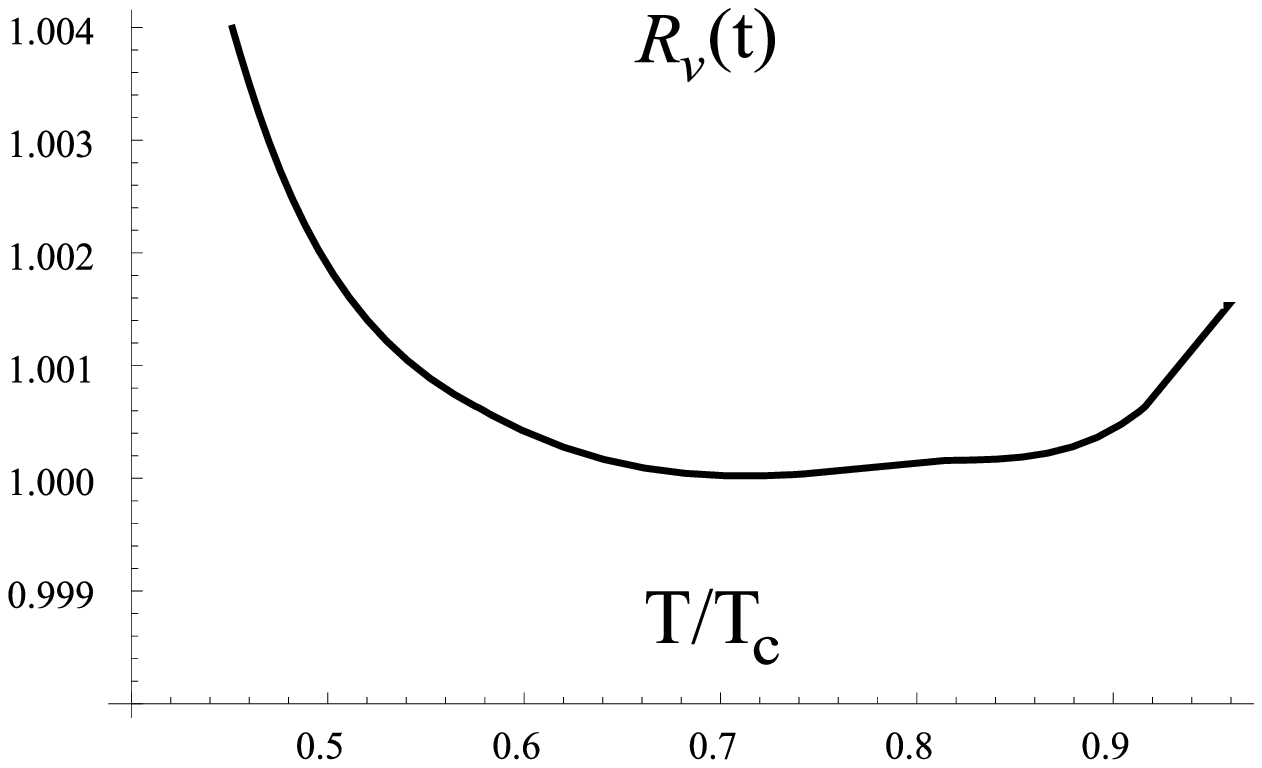}
\caption{Temperature dependence of the ratio $R_{v}  (t) =v^{(D_{2} O)}(t) / v^{(H_{2} O)}(t)$ on the coexistence curves of the normal and heavy water \cite{nist69}.}\label{fig_rvheavywater}
\end{figure}

It seems to be natural to suppose that similar character of the rotational motion in normal and heavy water will also lead to the following estimate for the ratio of their specific heats:
\[{\frac{{\vert C_{\upsilon}^{(H_{2}O)}(t) - C_{\upsilon}^{(D_{2}O)}(t)\vert}} {{C_{\upsilon}^{(H_{2}O)} (t)}}} \approx \mu . \]
This ratio is really in correspondence with experimental data, presented in Fig.~\ref{fig_cvh2odo}.
However, here we would like to focus our attention on the practically constant value of the difference $C_{\upsilon}^{(D_{2} O)} (t) - C_{\upsilon}^{(H_{2} O)} (t)$ in the temperature interval $0.8 < t < 0.95$. It corresponds to the fluctuation region, in which the normal and heavy waters are strongly dimerized. As we see, in this region $C_{\upsilon}^{(D_{2} O)} (t) - C_{\upsilon}^{(H_{2} O)} (t) \approx \frac{1}{2}.$ This effect finds the natural explanation if we assume that the inner rotation in dimers $(D_{2} O)_{2} $ is possible (see Fig.~\ref{fig_dimer_rotation}).
The activation of this type of thermal motion in heavy water is justified by greater length of D-bonds in comparison with H-bonds \cite{water_normalheavy_prl2008} and low energetic barriers \cite{dimers_water_shipman_jpc1974}. This circumstance becomes evident from the comparison of the specific volumes for normal and heavy waters (Fig.~\ref{fig_rvheavywater}).
The values of $R_{\upsilon}  (t)$ near the triple points, $t_{tr} \approx 0.42$, are mainly caused by different lengths of H- and D-bonds, since the numbers of them per molecule in normal and heavy water are practically the same \cite{water_dimermc_jcp2007}. From Fig.~\ref{fig_rvheavywater} it follows that the length of a D-bond is greater than one for an H-bond approximately $1.5\%$.
\section{Conclusion}
The main attention in this paper is focused on the physical nature of the temperature dependence of main thermodynamic properties of water: density, heat of evaporation and entropy as well as specific heat everywhere on the coexistence curve. All temperature interval of the two-phase vapor-liquid states is considered.

Analyzing the behavior of the density and the heat of evaporation of water we have established that these quantities have argon-like behavior everywhere excepting the fluctuation region near the critical point. It means that such a character of the temperature dependence is formed by averaged argon-like inter-particle potential, which arises in consequence of the rotation of water molecules. H-bonds bring in the specificity in the rotational motion of molecules, however their influence is not crucial. This circumstance is justified by that fact that the shear viscosities of water and argon have the same order of magnitude. The situation is considerably changed only for supercooled states of water and near its critical point. In the present paper we touch some details only for the near critical behavior.

Here the volume occupied by two water molecules becomes greater than that volume, which corresponds to a rotating dimer. As a result, the favorable conditions for the dimerization are created. Applying the method of chemical equilibrium it is shown that the degree of dimerization exceeds 0.9. Due to dimerization the characteristic ratio $R_{\upsilon}  (t)$ for specific volumes of water and argon increments approximately 15\% in the vicinity of their critical points. It shown that the ensemble of dimers with small admixture of monomers allows to reproduce successfully the location of the critical point of water. It is also responsible for the correct relations between critical amplitudes.

At elongation from the fluctuation region dimers destroy and short-living linear chains of water molecules are formed. Here the thermodynamic properties of water are determined by rotating monomers (of course, the rotation of monomers is not free). Due to formation of short living H-bonds the effective proper volume, corresponding a monomer, becomes to be dependent on the temperature and density, or pressure. Including the dependence of the proper molecular volume on the pressure to the van der Waals or Carnahan-Starling equations of state leads to the very nontrivial consequences. Among them we mark out 1) the non-monotone temperature dependence of the diameter of entropy and 2) the differences in the values of the specific heats for vapor and liquid phases, calculated with the help of the modified van der Waals or Carnahan-Starling equations of states. Here it is appropriate to note that the specific heats for coexisting phases are identical if the proper molecular volume is constant. Besides, the change of the sign of the entropy diameter at the lower limit for the fluctuation region $t_{u}  \approx 0.8$ is not also occasional. It is a strong argument in the favor that this root of the entropy diameter is connected with the discrepancy of the mean field and experimental values of entropy at the critical point.

The diameter of entropy changes also its sign at the lower temperature $t_{l} \approx 0.5$. The physical prerequisite for this root of the entropy diameter is the following: near $t_{l} $ the ensemble of linear molecular chains begins to form the spatially arranged the H-bond network. The rotation of molecules becomes more difficult and the character of the density dependence of the proper molecular volume essentially changes.

The very nontrivial manifestation of the isotopic effects is observed at the comparison of the temperature dependencies for the diameter of entropy and the specific heat of the normal and heavy water. Here we pay attention on the characteristic values of differences in the positions of the upper and lower dimensionless temperature for the normal and heavy water. These differences are close to each other and they practically coincide with dimensionless parameter describing the discrepancy in the rotations of dimers arising in vapor and liquid phases of normal and heavy water. Another remarkable fact is the difference of the specific heats for the normal and heavy waters in the fluctuation temperature interval $0.85 < t < 0.98$. In accordance with our reasons it arises owing to switching on the internal rotation for dimers in the heavy water.

The small differences in the values of the crystallization and critical temperatures for them is connected with some distinction of their averaged intermolecular potentials. In consequence of the strong dimerization near the critical point the normal and heavy water form the separated class from the point of view of the principle of corresponding states. Besides water, many alcohols belong probably to this class of corresponding states.

\appendix
\section{The degree of dimerization of water near the critical region (mean field analysis)}\label{sec_app_chemeq}
The equation for the chemical equilibrium between dimers and
monomers is:
\begin{equation}\label{chemeq}
  \f{A}{(1-A)^2} = 2n_0 \,K(T)\,\exp\left(\,\f{2\mu^{(ex)}_{1} - \mu_{2}^{(ex)}}{T}\,\right)\,,
\end{equation}
where $\mu_i^{(ex)}$ are the excess chemical
potentials (with respect to the ideal gas term) of the monomers and dimers correspondingly which are determined from the excessive part of the free energy corresponding to the EOS \eqref{bertlo_eos}.
The quantity $K(T)$ is the constant of chemical equilibrium we can approximate it by simple expression \cite{book_ll5}:
\begin{equation}\label{k}
  K(T) = \exp\left(\,\f{E_d}{T} \,\right)\,.
\end{equation}
The dissociation energy $E_d$ corresponds to the energy of H-bond which in $k_B\, T_c$
units is of order $E_{H} = 3\div 4\, k_B T_c$ \cite{book_water_zacepina,water_malenkov}.
Such a large value of the H-bond energy allows us to conclude that near critical water is dimerized to a great extent \cite{water_dimers_raman_jcp1998, water_dimers_infrared_jcp2003}. In particular it agrees with the results of the numerical simulations \cite{water_dimersimul_prl2000}.

In order to calculate the dimerization degree at the critical point we use the model Eq.~\eqref{bertlo_eos} with the dissociation considered as the perturbation to the completely dimerized state. In this way we obtain the critical temperature in the linear approximation on $e^{-E_d/2}<1$:
\[T_c = 1+\f{ {\tilde{b}_1}^{2}-6\,\tilde{b}_1\,\tilde{a}_{12}-4\,\tilde{a}_{12}+9\,{
\tilde{a}_{12}}^{2}}{12}\, e^{-E_d/2 -\f{1}{2}\,\tilde{b}_1+\f{9}{4}\,\tilde{a}_{12}-\frac {7}{8}} + \ldots\,\,\,.
\]
where
\begin{align}\label{bertlo_tilde_ab}
\tilde{b}_1 = \left(\,\f{\sigma_1}{\sigma_2}\,\right)^3\,,\quad \tilde{a}_{12} = 8\,\lambda \,
\f{\left(\,\sigma_1/\sigma_2\,\right)^2}{\left(\,1+\f{\sigma_1}{\sigma_2}\,\right)^3}\,,
\end{align}
The corresponding value for $Z_c$ is:
\begin{equation}\label{zc_assoc}
  Z_c = Z^{(0)}_c\left(\,1-A_c/2\,\right)\,.
\end{equation}
The dependence of $Z_c$ on the relevant parameters is shown on Fig.~\ref{fig_zc}.
E.g. at $\tilde a_{12}\approx 0.2$, $\tilde{b}_1 = 0.2$
and $E_d/T_c = 3.5$  it takes the value $Z_c\approx 0.2$.
\begin{figure}[hbt!]
  \includegraphics[scale=0.5]{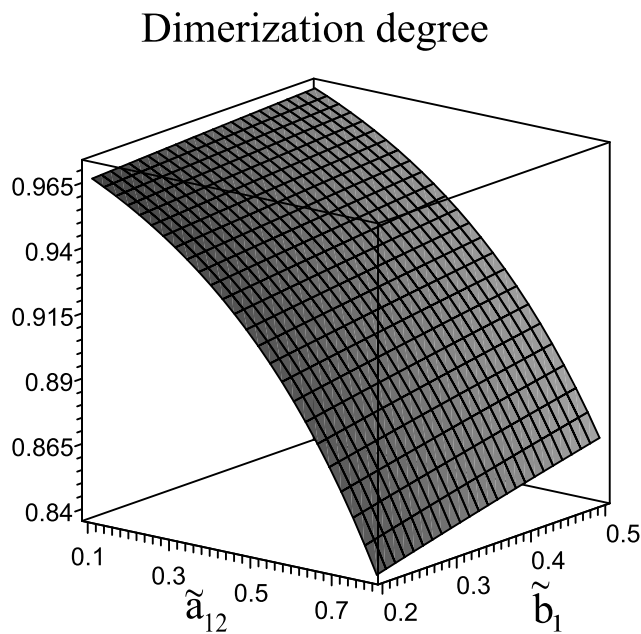}\\
  \caption{Dimerization degree as the function of the parameters of monomer-dimerization interaction}\label{fig_dimerization}
\end{figure}
In the vicinity of the critical point assuming that
$e^{-\f{E_d}{T_c}}\ll 1\,,$ from \eqref{chemeq}
we get the degree of the dimerization:
\begin{equation}\label{dimerizationdegree}
  A_c = 1-\f{1}{3}\, e^{-E_d/2 -\f{1}{2}\,\tilde{b}_1+\f{9}{4}\,\tilde{a}_{12}-\frac {7}{8}}+O \left( {\epsilon}^{2} \right)\,,
\end{equation}
where $\lambda = \left(\,\f{d_{1}\sigma_2}{d_{2}\sigma_1}\,\right)^2$ is the parameter which determines the difference in charge distribution in the monomer and the dimer. The dependence of the dimerization factor on the parameters of the monomer-dimerization interaction is shown on Fig.~\ref{fig_dimerization}.

Thus in the critical region the degree of dimerization $A$ at $\tilde a_{12}\approx 0.4$, which corresponds to $d_1/d_2\approx 0.5\,, \sigma_1/\sigma_2\approx 0.7$, is approximately $0.9$. This conforms with the thermodynamic estimates given in Section~\ref{sec_intro} and the results of works \cite{water_dimer_us_nato2007}.
\newpage

\end{document}